\documentclass[twocolumn,showpacs,preprintnumbers,amsmath,amssymb,groupaddress]{revtex4}
\usepackage{bbm}
\usepackage{amsfonts}

   \usepackage{CJK}
 \usepackage{amsmath,cleveref}
 \usepackage{graphicx}
\begin{document}

\begin{CJK*}{GBK}{}

%\preprint{CREAM/MS-2004-01-PRA-GH-01}

\title{Transition between self-focusing and self-defocusing in nonlocally nonlinear media
}% Force line breaks with \\
\author{Guo Liang$^{1,5,*}$,Weiyi Hong$^{1,*}$,Yahong Hu$^{2,}$\footnote{The first two authors have contributed equally to this work, with the first one mainly to analytical operations and the second one mainly to numerical simulations. The third author has contributed to analytical solution of the dark soliton.\\
$^\dag$Electronic address: guoq@scnu.edu.cn}}
%\thanks{The first two authors have contributed equally to this work, with the first one mainly to analytical operations and the second one mainly to numerical simulations. The third author has contributed to analytical solution of the dark soliton.\\
%$^\dag$Electronic address: guoq@scnu.edu.cn}
\author{Jing Wang$^1$,Zhuo Wang$^1$,Yingbing Li$^1$,Qi Guo$^{1,\dag}$,Wei Hu$^1$,Senyue Lou$^3$}
\author{Demetrios N. Christodoulides$^4$}
%\affiliation{Department of Physics, Ningbo University, Ningbo 315211, P. R. China}
\affiliation{$^1$Guangdong Provincial Key Laboratory of Nanophotonic Functional Materials and Devices, South
China Normal University, Guangzhou 510631, P. R. China\\
$^2$Department of Mathematics, Zhejiang Lishui
University, Lishui 323000, P. R. China\\
$^3$Department of Physics, Ningbo University, Ningbo 315211, P. R. China\\
$^4$CREOL/College of Optics, University of Central Florida, Orlando, Florida 32816, USA\\
$^5$School of Physics and Electrical Information, Shangqiu Normal University, Shangqiu 476000, P. R. China
}

\date{\today}% It is always \today, today,

\begin{abstract}
We reveal the relevance between the nonlocality and the focusing/defocusing states in nonlocally nonlinear media, and predict a novel phenomenon that the self-focusing/self-defocusing property of the optical beam in the nonlocally nonlinear medium with a sine-oscillation response function depends on its degree of nonlocality. The transition from the focusing nonlinearity to the defocusing nonlinearity of the nonlinear refractive index will happen when the degree of nonlocality of the system goes cross a critical value, and vise verse. Bright and dark soliton solutions are obtained, respectively, in the focusing state and in the defocusing state, and their stabilities are also discussed. It is mentioned that such a phenomenon might be experimentally realized in the nematic liquid crystal with negative dielectric anisotropy or in the quadratic nonlinear medium.
\end{abstract}


%\keywords{Use showkeys class option if keyword display desired}%
\maketitle
\end{CJK*}
The optical Kerr effect(OKE)~\cite{Shen-book-1984,Boyd-book-no,guo-book-2015}, as one of the most important effects in nonlinear optics, is a fundamental and widespread phenomenon in the nonlinear interactions of light with materials, such as semiconductors~\cite{Sheik-Bahae-prl-1990}, polymers~\cite{Blau-prl-1991}, liquid crystals~\cite{conti-prl-2003,assanto-book-2013}, soft matters~\cite{Conti-prl-2005}, photorefractive~\cite{Segev-prl-1992} and thermal~\cite{Rotschild-prl-2005,Ghofraniha-prl-2007,Dreischuh-prl-2006} media. The equivalent OKE can also be found in optical quadratic nonlinear processes~\cite{Esbensen-pra-2012,Nikolov-pre-2003,Buryak-pla-1995}, and the other physical systems, such as Bose-Einstein condensates~\cite{Pedri-prl-2005}, quantum electron plasmas~\cite{Shukla-prl-2006}, and even on the surface of water~\cite{Chabchoub-prl-2013}. The OKE refers to the light-intensity dependence of the refractive index $n$,
%the intensity-dependent refractive index
 that is, $n=n_0+N,$ where $n_0$ is its linear part and $N$ is the light-induced nonlinear refractive index (NRI).
 %The OKE is also called the self-action effects in the sence that ... .
 Optical solitons~\cite{kivshar-book-2003,guo-book-2015}
 % (light localization in space or time)
 are the main phenomena resulting from the OKE.

  The OKE is of two important intrinsic properties: the nonlocality and the focusing/defocusing.
The NRI exhibits generally the nonlocality both in space and time~\cite{guo-book-2015}. In consideration of the spatial nonlocality in bulk materials, the NRI can be expressed phenomenologically as~\cite{guo-book-2015,assanto-book-2013-chpter2} $N=n_2\int_{-\infty}^{+\infty}R(x-x')|E(x',z)|^2dx'$, where $n_2$ is the
 Kerr coefficient that is determined by material properties, the symmetric $R(x)$ is the response function of the media and $E$ the optical field.
If $R(x)$ becomes $\delta$-function, then $N=n_2|E|^2$, which is the well-known local OKE~\cite{Shen-book-1984,Boyd-book-no}; Otherwise, the nonlocality is non-negligible.
Systematic study on the nonlocality began with the work by Snyder and Mitchell~\cite{Snyder-science-1997}. Their work has attracted lots of attentions~\cite{assanto-book-2013,guo-book-2015,Peccianti-pr-2012,Krolikowski-JOB-2004,Guo-spie-2003,krolikowski-spie-2005}, and experiments about the spatial nonlocality have been carried out in nematic liquid
crystals~\cite{Conti-prl-2004}, lead glasses~\cite{Rotschild-prl-2005}, paraffin oils~\cite{Dreischuh-prl-2006}, and rhodamine aqueous solutions~\cite{Ghofraniha-prl-2007} for deeper and extensive investigations.
On the other hand, the focusing/defocusing of the OKE refers to the phenomenon that the optical beam propagating in the bulk medium with the homogeneous $n_0$ can focus or defocus itself by its induced NRI~\cite{Shen-book-1984,Boyd-book-no}. The material with $n_2>0$ or $n_2<0$ is called the self-focusing medium or the self-defocusing one, respectively~\cite{assanto-book-2013-chpter2}. It is commonly considered that
the focusing/defocusing property is determined only by the medium properties, and has nothing to do with the nonlocality.
In other words, the focusing/defocusing is irrelevant to the property of optical beams propagating in the medium.

In this letter, we will revisit the focusing/defocusing property of the media,
 and discover a dramatic relation between the focusing/defocusing and the nonlocality in the nonlocally nonlinear medium with a sine-oscillation
response function, which was introduced in the study of quadratic solitons, first obtained by Nikolov et al.,~\cite{Nikolov-pre-2003} and then mentioned in the other works~\cite{Esbensen-pra-2012,Wang-ol-2014}.
 By defining the focusing and defocusing states, we have found that in such a system there exist the focusing and defocusing states, and their inter-transition, which are related to the degree of nonlocality.
 Extensive discussions are also presented, including the bright and dark soliton solutions and their stability in the focusing and defocusing states, respectively.

\emph{Theoretical model}.---
We consider the propagation of the optical beam along $z$ axis in a nonlocally nonlinear medium described by the
system of equations for dimensionless complex optical field amplitude $\phi(x,z)$ and nonlinear refractive index $\Delta n(x,z)$ given by%~\cite{guo-book-2015}
\begin{subequations}\label{model-whole}
\begin{equation}
i\frac{\partial\phi}{\partial z}+\frac{1}{2}\frac{\partial^2\phi}{\partial x^2}+\Delta n\phi=0,\label{equation for optical beam}
\end{equation}
\begin{equation}
w_m^2\frac{d^2\Delta n}{d x^2}+\Delta n-s|\phi|^2=0,\label{equation for NRI}
\end{equation}
\end{subequations}
where $x~[\in(-\infty,\infty)]$ and $z~[\in[0,\infty)]$
 stand, respectively, for the transverse and longitudinal coordinates scaled to a beam width and the Rayleigh distance, and $w_m$ is the nonlinear characteristic length (NCL) of the system, and $s=\pm1$. When $w_m=0$
 (the local case), the NRI $\Delta n=s|\phi|^2$, and the system above is simplified into the well known nonlinear Schr\"{o}dinger equation
  $i {\partial_z\phi}+({1}/{2}){\partial_{x}^2\phi}+s|\phi|^2\phi=0$, which has the stable sech-form bright
  soliton for $s=1$ and the stable tanh-form dark soliton for $s=-1$, respectively~\cite{Agrawal-book-2001,guo-book-2015}. For the general case of nonzero $w_m$,
the NRI in Eq.~(\ref{equation for NRI}) can be obtained~\cite{Wang-ol-2014,Nikolov-pre-2003,Esbensen-pra-2012}
\begin{equation}\label{NRI with response function}
\Delta n(x,z)=s\int_{-\infty}^{\infty}R(x-x')|\phi(x',z)|^2dx',
\end{equation}
where $R(x)$ is the response function with the oscillatory form being full of the infinite space of $x$
\begin{eqnarray}\label{response-function}
 R(x)=\frac{1}{2w_m}\sin\left(\frac{|x|}{w_m}\right),
\end{eqnarray}
which was first obtained by Nikolov et al.~\cite{Nikolov-pre-2003}. The generalized degree of nonlocality (GDN) of the system is defined as $\sigma= w_m/w_r$,
where the beam width $w_r=({2\int_{-\infty}^{+\infty}x^2|\phi|^2dx/\int_{-\infty}^{+\infty}|\phi|^2dx})^{1/2}$~\cite{beamwidth-definition},
since the NCL $w_m$ determines the oscillation period of $R(x)$ and does not represent any more the scale occupied by $R(x)$ like the case of the
exponential-decay function~\cite{Rasmussen-pre-2005}. %, and is named the general degree of nonlocality here.
%This oscillatory
%response function was firstly obtained in Ref.~\cite{Nikolov-pre-2003}, where the parametric interaction in quadratic
%media was shown to be equivalent to nonlocal Kerr-type nonlinearities.
%In the Fourier $f$-domain, %(denoted by a tilde)
%the response function is given by a
%Lorentzian
%$\tilde{R}(f)=1/(1-w_m^2f^2),$ which has two poles on the real $f$-axis.
  %This response function for the case that $s=+1$ was first obtained in the quadratic nonlinear materials~\cite{Nikolov-pre-2003}, where it was showed that the parametric interaction in quadratic nonlinear materials is equivalent to the nonlocality in nonlocal nonlinear materials. This sine-oscillatory response function for the case that $s=-1$ can also describe the response to the light field in the nematic liquid crystals (NLC) with negative dielectric anisotropy~\cite{wang-arxiv}.

\begin{figure}[htb]
\centerline{\includegraphics[width=8cm]{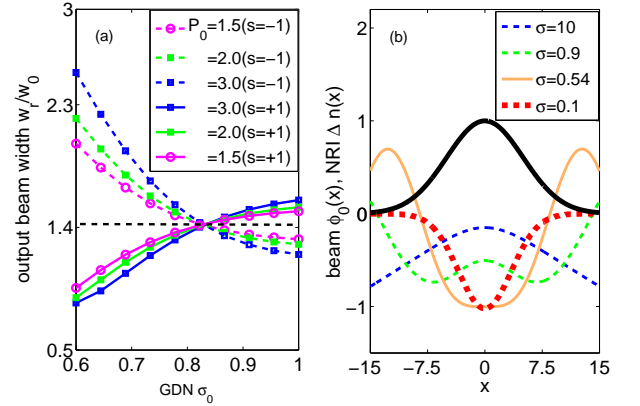}}
\caption{(color online) (a)  Normalized output beam widths $w_r/w_0$ at $z=1$ (one Rayleigh distance) as the function of $\sigma_0$, and the dashed black line presents the linear case where the analytic result is $w_r(z)/w_0=\sqrt{1+z^2}$~\cite{Haus-book-1984}. $w_m=2$ for all of the curves.
(b) Gaussian beam $\phi_0$ (thick solid black line, $w_0=5$) and its induced NRI for different $\sigma$ when $s=-1$.
The NRI for $\sigma=10$ is multiplied by 30.}\label{transition}
\end{figure}

%\newpage
\emph{Focusing/defocusing states and their inter-transition}.---
To glimpse at the focusing/defocusing property of the nonlinear system described by Eqs.~(\ref{model-whole}),
we simulate the propagation of an initial Gaussian beam with the input power $P_0$
\begin{equation}\label{Gaussian beam}
\phi_0(x)=\phi(x,z)|_{z=0}=\sqrt{\frac{P_0}{\sqrt{\pi}w_0}}\exp\left(-\frac{x^2}{2w_0^2}\right)
\end{equation}
in Eqs.~(\ref{equation for optical beam}) and (\ref{NRI with response function}), and show the typical results of the output beam width after propagating the distance $z=1$ for the different initial GDN $\sigma_0(=w_m/w_0)$ (the different $w_0$ and the fixed $w_m$) in Fig.~\ref{transition} (a). As can been observed in the figure, for a given $P_0$, the output beam widths will be larger or smaller, in the smaller or bigger sides of the $\sigma$-coordinate respectively, than linear case when $s=-1$; the inverse results will be got when $s=1$. The higher $P_0$, the stronger the effect. It is well known that~\cite{Boyd-book-no,Shen-book-1984} the optical beam sampling the defocusing nonlinearity expands faster than that of the linear case; and the focusing case follows an opposite trend.
Obviously, the focusing/defocusing nonlinearity sampled by the optical beams depends on its GDN $\sigma$ dramatically for the nonlocal system~(\ref{model-whole}). For the case that $s=-1$, the transition from the self-focusing
 to the self-defocusing will happen when $\sigma$ goes down cross the critical points that are the same and nothing to do with $P_0$, and vice verse. The case that $s=+1$ is on the contrary.

The phenomenon observed above can be well understood through the relationship between the variations of the light intensity and its induced NRI distributions. We define
 the focusing/defocusing state of the NRI by the variation of the NRI distribution against that of the light intensity on the transverse perpendicular to the propagation direction $z$. The NRI has two states: focusing and defocusing. For the focusing state, the NRI changes uniformly with the intensity, that is, the NRI increases as the intensity increases, and vice versa; The defocusing state is on the contrary.
%The state that the NRI changing uniformly with the intensity is the focusing state, that is, the NRI increases as the intensity increases, and vice versa.  The defocusing state is on the contrary.
 %For a bell-shaped beam, the focusing and defocusing states exist under conditions of
% negative and positive $d^2\Delta n/dx^2|_{x=0}$, respectively.
 According to the definition, the focusing/defocusing states depend on the convexity-concavity of the NRI curves, and the sufficient condition for the realization of the transition between the focusing/defocusing states is that $d^2\Delta n/dx^2|_{x=0}=0$.
 %the second order derivative of the NRI at the symmetry point must be zero.
 For the local OKE~\cite{Shen-book-1984,Boyd-book-no} and the nonlocal OKE with non-oscillatory $R(x)$~\cite{guo-book-2015,Krolikowski-JOB-2004,Rasmussen-pre-2005,assanto-book-2013,Rotschild-prl-2005}, for example, the exponential-decay function~\cite{Rasmussen-pre-2005,guo-book-2015}, the focusing/defocusing states are only determined by sgn($s$)
 (the Kerr coefficient $n_2$ in the actual physical system). The focusing state appears when $s=1$ ($n_2>0$), and the defocusing state does when $s=-1$ ($n_2<0$).
The focusing/defocusing property of those two cases is determined only by the medium properties, and has nothing to do with the nonlocality, that is, irrelevant to the property of optical beams propagating in the medium.
 No transition between them can happen in both cases because $d^2\Delta n/dx^2|_{x=0}\neq0$.
%The positive Kerr coefficient determines the focusing state of the media while the negative one corresponds to the defocusing state.
The system ~(\ref{model-whole}) with a sine-oscillation $R(x)$, however, can realize the transition because $d^2\Delta n/dx^2|_{x=0}=0$ for some critical points of the GDN. Since the NRI depends on the specific profile of the
beams, by taking the Gaussian beam given in Eq.~(\ref{Gaussian beam}), we obtain
%The NRI near $x=0$ is inverted bell-shaped when $d^2\Delta n/dx^2|_{x=0}>0$, and is bell-shaped when $d^2\Delta n/dx^2|_{x=0}<0$.
%Substituting Eq.~(\ref{Gaussian beam}) into Eq.~(\ref{NRI with response function}), and taking second-order derivative of $\Delta n$ with respect to $x$ yields
$\left.d^2\Delta n/dx^2\right|_{x=0}=-sP_0\left[F\left(1/2\sigma\right)-\sigma\right]/\sqrt{\pi}w_m^3$, where $F(x)=\exp(-x^2)\int_0^{x}\exp(t^2)dt$ is the Dawson function, and a critical GDN
$\sigma_c=0.54$.
%\begin{figure}[htb]
%\centerline{\includegraphics[width=5cm]{figure-critical.eps}}
%\caption{(color online) Profiles of NRI for different GDNs near $\sigma_c$ for the case of $s=-1$. The profile of the beam given by Eq.~(\ref{Gaussian beam}) with $w_0=5$ is presented in the solid magenta line.} %changeless. }
%\label{critical}
%\end{figure}
Fig.~\ref{transition} (b) shows the transition process for $s=-1$ by giving the different curves of the NRI for different $\sigma$.
% is the profiles of NRI for different GDNs.
%When $\sigma=\sigma_c$, the central part of $\Delta n(x)$ is flat.(such as $\sigma=0.1$)
For smaller $\sigma$ such that $\sigma<\sigma_c$, the NRI is defocusing, and the beam samples the defocusing index, then will be self-defocused, as can be observed in Fig.~\ref{transition} (a).
%When $\sigma=\sigma_c$, the central part of $\Delta n(x)$ is flat.
When $\sigma$ exceeds $\sigma_c$, the focusing state of the NRI appears in the center, and both the range and the amplitude of the bell-shaped NRI increases as $\sigma$ increases. When $\sigma$ is slightly larger than $\sigma_c$, moreover,
the NRI is partially focusing and partially defocusing such that the central part of the beam samples the focusing index and its edges ``sees'' the defocusing one.
%$ is slightly larger than $\sigma_c$.
The optical beam as a whole will continue to exhibit the self-defocusing behavior until the effect of the focusing-index dominates.
  %between the focusing state and defocusing state.
   This can explain why $\sigma_c$ is smaller than the critical points in Fig.~\ref{transition} (a).
    The exact inverse is the case that $s=1$ .
 The transition from the self-focusing to self-defocusing of the optical beam will happen when the GDN $\sigma_0$ goes up across a critical value.
%%   The inverse happens for $s=1$ that
%%   the NRI is, respectively, in the focusing state and the mixed state of the focusing/defocusing
%   when $\sigma$ .

   %, when $\sigma$ is slightly larger that $\sigma_c$, the defocusing state dominates over the focusing state, then the optical beam is still be defocused.

\begin{figure}[htb]
\centerline{\includegraphics[width=7cm]{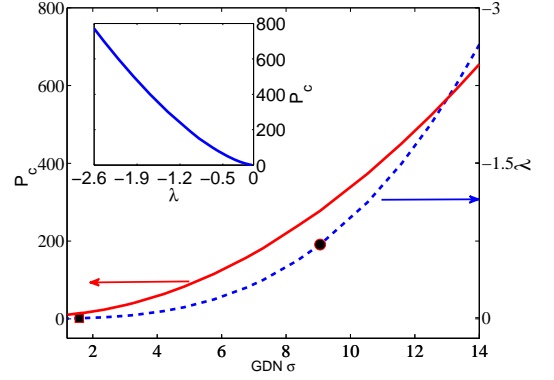}}
\caption{(color online) Critical power $P_c$ and
the soliton propagation constant $\lambda$ versus the GDN $\sigma$. The inset shows $P_c$ versus $\lambda$. All of the results are obtained for $w_m=10$.}\label{solitonspectrum}
\end{figure}

\begin{figure}[htb]
\centerline{\includegraphics[width=9cm]{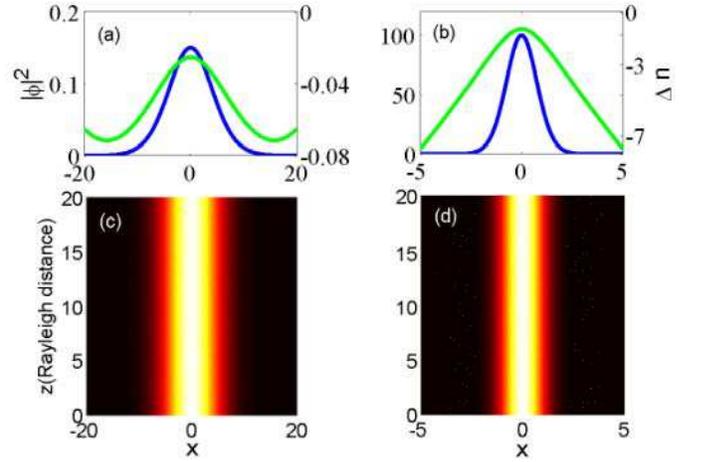}}
\caption{(color online) The solitons (solid blue lines) for the small $\sigma(=1.58)$ (a) and the large $\sigma(=9.06)$ (b), which are corresponding, respectively, to the filled square and the filled circle in Fig.~\ref{solitonspectrum}, and solid green lines represent the corresponding NRI. The numerial propagations of the solitons with $5\%$ random noises are shown in (c) and (d), respectively.}\label{compare}
\end{figure}

\emph{Bright solitons}.---In the higher range of the GDN $\sigma$ ($\sigma>\sigma_c$) for the case that $s=-1$, optical beams sample the focusing index, and bright solitons can form when
the nonlinear effects balance the diffractive effects precisely.
By the imaginary-time method~\cite{Chiofalo-pre-2000}, the numerical soliton solutions, $\phi=u(x)\exp(i\lambda z)$, of Eq~(\ref{equation for optical beam}) plus (\ref{NRI with response function}) do be obtained in the range $\sigma>\sigma_{bsc}(\sigma_{bsc}=1.21)$! No soliton solutions can be found in the lower range that $\sigma<\sigma_{bsc}$.
%In the strongly nonlocal case ($\sigma\gg1$), the approximate analytical soliton solutions can be obtained~\cite{Supplemental Material}.
%Bright solitons exist when $\sigma>1.212$,
Shown in Figs.~\ref{solitonspectrum} and~\ref{compare} are the $\sigma$-spectrums of the critical power of the soliton $P_c(=\int u^2dx)$ and the soliton propagation constant $\lambda$ (that is, the dependences of $P_c$ and $\lambda$ upon $\sigma$), and the distributions of the soltion, respectively. All of
the bright solitons are shown to be stable~\cite{Supplemental Material} by the linear stability analysis~\cite{Xu-ol-2005}, which also be confirmed by the simulations, as shown in Fig.\ref{compare}. The bright solitons of the system described by Eqs.~(\ref{model-whole}) [equivalently Eq.~(\ref{equation for optical beam})
plus (\ref{NRI with response function})] with $s=-1$ have two abnormal properties. First, the soliton propagation constants $\lambda$
are negative, which results from the negative $\Delta n(x)$~\cite{explain-lambda-negative}. For the bright solitons obtained before in the local nonlinear media \cite{kivshar-book-2003,guo-book-2015} and the nonlocallly
 nonlinear media with the non-oscillatory response function~\cite{assanto-book-2013,guo-book-2015,Peccianti-pr-2012,Krolikowski-JOB-2004,Guo-spie-2003,krolikowski-spie-2005,Guo-pre-2004,Shou-ol-2011}, however, their propagation constants are all positive.
 Second, the slope of the $P_c(\lambda)$
is negative, as shown in Fig.~\ref{solitonspectrum}. Therefore, their stability criterion obeys an inverted Vakhitov-Kolokolov stability
criterion~\cite{Sakaguchi-pra-2010}.

\begin{figure}[htb]
\centerline{\includegraphics[width=8cm]{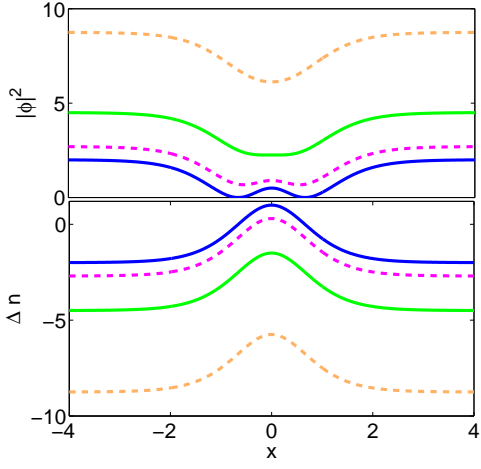}}
\caption{(color online) The dark soliton solutions (figure up) and the corresponding NRI (figure down). $\sigma$ is $0.31,~0.29~(w_m/w_0=\sqrt{1/5}),~0.24,$ and $0.18~(w_m/w_0=1/4)$  from the bottom to the
top (up) and from the top to the bottom  (down), respectively.}\label{darksoliton}
\end{figure}

\emph{Dark solitons}.---In the defocusing side of $\sigma$ ($\sigma<\sigma_c$) for $s=-1$,
the optical beam samples the defocusing index. The exact dark soliton solution is found to exist in the condition that $w_m/w_0\leq1/2$~\cite{Hu-ctp}
%\begin{widetext}
%\begin{equation}\label{qsolution}
%\phi(\xi)=\frac{\sqrt{2}w_m}{2w_0^2}\sqrt{36\tanh^4\left(\xi\right)+\left(\frac{6w_0^2}{w_m^2}-48\right)\tanh^2\left(\xi\right)+4-\frac{(4w_m^2/w_0^2-1)}{w_m^4/w_0^4}}\exp(i\varphi),
%\end{equation}
%\end{widetext}
\begin{equation}\label{qsolution}
\phi(\xi)=\gamma_1\sqrt{\tanh^4(\xi)+\gamma_2\tanh^2\left(\xi\right)+\gamma_3}\exp(i\varphi),
\end{equation}
and %the induced NRI is
$\Delta n=4w_m^2/w_0^4+2/w_0^2-1/2w_m^2-3\tanh^2(\xi)/w_0^2,$
where the phase
$
\varphi=\theta+\left(4w_m^2/w_0^4-v^2/2-3/4w_m^2\right)\xi,
$
$
\theta=\theta_0+vx+C(T_++T_-)+C\xi/(\kappa_--1)(\kappa_+-1)
$,
$\xi=(x-vz)/w_0$, $\kappa_{\pm}=2/3-(1\pm\sqrt{48w_m^4/w_0^4-3})/(12w_m^2/w_0^2)$, $T_{\pm}=\arctan[\tan(\xi)/\sqrt{\kappa_\pm}]/\sqrt{\kappa_\pm}(\kappa_\pm-1)(\kappa_\pm-\kappa_\mp)$, $C^2=(\kappa_+-1)^2(\kappa_--1)^2\left(w_0^2/2w_m^2-2\right)$, $\gamma_1=3\sqrt{2}w_m/w_0^2$, $\gamma_2=(w_0^2/w_m^2-8)/6$, and $\gamma_3=(4w_m^4/w_0^4-4w_m^2/w_0^2+1)/(36w_m^4/w_0^4)$. The parameter $v$ denotes the tangent of an incident angle of the dark soliton. Without loss of generality, the case of the normal incidence ($v=0$) is given in Fig~\ref{darksoliton}. By defining the width of the dark soliton as $w_r=[{2\int_{-\infty}^{+\infty}x^2(|\phi_{\infty}|^2-|\phi|^2)dx/\int_{-\infty}^{+\infty}(|\phi_{\infty}|^2-|\phi|^2)dx}]^{1/2}$ with $|\phi_{\infty}|$ being the background amplitude, we have its GND $\sigma=(\sqrt{2}w_m/w_0)/\sqrt{\pi^2/3+8(w_m/w_0)^2}$, and then find that the $\sigma$-range for the exist of the dark soliton solution is $\sigma\leq\sigma_{dsc}(= 0.31)$. When $\sigma=0.31 (w_m/w_0=1/2)$, the dark soliton [Eq.(\ref{qsolution})] is deduced to the bi-black soliton $|\phi|=\left|3\text{sech}^2\left(x/2w_m\right)-2\right|/2\sqrt{2}w_m$, and has two
zero dips, which was obtained in Ref.~\cite{Buryak-pla-1995}.
When $\sigma\leq0.24 (w_m/w_0\leq1/2\sqrt{2})$, the dark soliton becomes the single gray soliton with one nonzero dip.
 The dark soliton in the range that $0.24<\sigma<0.31$ is the bi-gray soliton with two nonzero dips.

Modulation instability (MI) of system ~(\ref{model-whole}) has been discussed, and the main result is that
the MI always occur except for the case that $s=-1$ and $w_m=0$~\cite{Supplemental Material}. Therefore, the dark soliton solutions (\ref{qsolution}) above are unstable when $w_m\neq0$ due to the MI.
When $w_m=0$, however, the solution (\ref{qsolution}) does not work because of the term $w_m$ in denominator. As a result, the system described by Eq.~(\ref{model-whole}) has two different kinds of dark soliton solutions in the range that $\sigma<0.31$, which are the unstable soliton solutions (\ref{qsolution}) when $w_m\neq0$ and the stable solitons~\cite{guo-book-2015,Agrawal-book-2001} when $w_m=0$, respectively.

\emph{The situation that $s=1$}.---
As discussed above, optical beams sample, respectively, the defocusing and focusing indexes in the higher and lower ranges of the GDN for such a situation. The case that $s=1$ and the large GDN was discussed in Ref.~\cite{Esbensen-pra-2012}, and no single-hump bright soliton was found, which is consistent with the phenomenon we discovered. When the boundary is involved, however, the single-hump bright soliton can sometimes exist for the case because the boundary can re-distribute the NRI of the system~\eqref{model-whole} such that $\Delta n$ can become focusing in the center region for some suitable conditions~\cite{Wang-ol-2014}.
The dark solitons in the defocusing side ($\sigma>\sigma_c$),
 %of $\sigma$ ($\sigma>\sigma_c$)
 even if they exist, are unstable due to the existing MI. Therefore, we only take the bright soliton in the focusing side ($\sigma<\sigma_c$) into consideration. Because the scope of optical beams is much larger than $w_m$ for the small GND, to numerically search the soliton solution requires extremely high-precision sampling of the NRI, which leads to a huge need for computing sources.
  %even for the 1+1-dimensional problem.
   Due to the computer source limit, the (single-hump) bright soliton has not been found so far. We cannot be sure, therefore, whether the bright soliton exist in the small side ($\sigma<\sigma_c$) of the $\sigma$-coordinate, although the necessary condition is satisfied. What we can be sure, however, is there exists the sech-form bright soliton in Eqs.~\eqref{model-whole} in the smallest end that $w_m=0$~\cite{guo-book-2015,Agrawal-book-2001}.

\emph{Possibilities of experimental realization}.---
The
 propagation of the (1+1)-dimensional optical beam $\Phi$ in the nematic liquid crystals (NLC) with negative dielectric anisotropy~\cite{wang-arxiv-2015} can be described by the model~(\ref{model-whole}) for $s=-1$. Following the procedure in Ref.~\cite{wang-arxiv-2015}, a set of equations satisfied by $\Phi$ (the optical beam) and $\Psi$ (the light-induced perturbation of the tilt angle of NLC molecules) can be obtained
\begin{subequations}\label{model-nlc-whole}
\begin{equation}
i\frac{\partial \Phi}{\partial Z}+\frac{1}{2k}\nabla_{XY}^2\Phi+k_0N_L \Phi=0,\label{simplified beam equation}
\end{equation}
\begin{equation}
 W_{mL}^2\nabla^{2}_{XY}N_L+N_L-n_2^L|\Phi|^{2}=0,\label{simplified NRI equation}
\end{equation}
\end{subequations}
%
%with $N_L$ being the optically induced perturbation,
where $N_L=\epsilon_a^{op}\sin(2\theta_0)\Psi/2n_0$, $
W_{mL}=\left\{2\theta_0K/\epsilon_0|\epsilon_a^{rf}|\sin(2\theta_0)[1-2\theta_0\cot(2\theta_0)]\right\}^{1/2}/E_{rf}$,
 $n_2^L=(\epsilon_a^{op})^2\theta_0\sin(2\theta_0)/4n_0\epsilon_a^{rf}E_{rf}^2[1-2\theta_0\cot(2\theta_0)]$, $k=n_0k_0$, $k_0=\omega/c$, and
 the other symbols not specified here are the same with those in Ref.~\cite{wang-arxiv-2015}. A family of Equations (\ref{model-nlc-whole}) are the standard
 form of the equation for the Kerr-type nonlocal nonlinear process~\cite{guo-book-2015}, and $ N_L$, $n_{2}^L~(n_{2}^L<0, \text{because}~\epsilon_a^{rf}<0)$, and $W_{mL}$ are, respectively,
 the NRI, the Kerr coefficient, and the NCL for the NLC with negative dielectric anisotropy.
Introducing the dimensionless transform $\phi=\Phi/\Phi_0$, $\Delta n=N_L/N_{L0}$, $x(y)=X(Y)/W_0$, $z=Z/kW_0^2$, and $w_m=W_{mL}/W_0$,
where $\Phi_0=\sqrt{n_0/|n_2^L|}/kW_0$ and $N_{L0}=1/k_0^2W_0^2n_0$, the
system can be expressed in the dimensionless form as
$ i\partial_z \phi+(1/2)\nabla_{xy}^2\phi+\Delta n\phi=0$ and
   $ w_m^2\nabla_{xy}^2\Delta n+\Delta n+|\varphi|^2=0$,
the $(1+1)-D$ case of which corresponds to Eqs.(\ref{model-whole}) with $s=-1$. As a matter of fact, the stable bright soliton has been experimentally observed in such a configuration~\cite{wang-arxiv-2015}.
Since $W_{mL}$ can be controlled by the bias voltage~\cite{hu-apl-2006}, we might expect to realize the switching the focusing/defocusing states by the bias voltage in the NLC with negative dielectric anisotropy.

The second candidate to experimentally observe the new phenomenon discovered might be the interaction process between the fundamental wave $E_1(X,Z)$ and the second-harmonic wave $E_2(X,Z)$ in quadratic nonlinear media. For the type-I phase matching, the process is described by~\cite{Esbensen-pra-2012},
$i{\partial_Z E_1}+\gamma_1{\partial^2_{X}E_1}+\chi_1E_2E^{*}_1\exp{(-i\Delta k Z)}=0$,
$i{\partial_Z E_2}+\gamma_2{\partial^2_{X}E_2}+\chi_2E_1^2\exp{(i\Delta k Z)}=0$,
where all symbols are same with those in Ref.~\cite{Esbensen-pra-2012}. As suggested in Ref.~\cite{Esbensen-pra-2012,Nikolov-pre-2003}, it can be assumed that $E_2=e_2(X,Z)\exp(i\Delta k Z)$, and $\partial_Ze_2$ can be neglected comparing with the terms left~\cite{Esbensen-pra-2012,Krolikowski-JOB-2004}. In this way, we can have the standard equations
%i\partial E_1/\partial Z+\gamma_1\partial^2E_1/\partial X^2+\chi_1e_2E^{*}_1=0,
%\gamma_2\partial^2e_2/\partial X^2-\Delta k e_2+\chi_2E_1^2=0,
%$
$i{\partial_Z E_1}+s_1|\gamma_1|{\partial_X^2E_1}+k_0 N_Q E^{*}_1=0,$
%\end{equation}
%\begin{equation}
$s_2W_{mQ}^2{\partial_X^2 N_Q}- N_Q+n_{2}^Q E_1^2=0,$
%\end{equation}\end{subequations}
where $ N_Q=\chi_1 e_2/k_0$, $n_{2}^Q={\chi_2\chi_1}/{k_0\Delta k}$, $W_{mQ}=\sqrt{\left|\gamma_2/\Delta k\right|}$, $s_1=\mathrm{sgn}(\gamma_1)$, and $s_2=\mathrm{sgn}(\Delta k\gamma_2)$. $ N_Q$, $n_{2}^Q$, and $W_{mQ}$ are the equivalent NRI, the equivalent Kerr coefficient, and the NCL for quadratic nonlinear process, respectively.
By the dimensionless trasnform %$\phi$ and $\Delta n$, as well as the dimensionless variables $x$ and $z$,
$
\phi=E_1/E_{10},\Delta n=N_Q/N_{Q0},x=X/W_0,z=2|\gamma_1|Z/W_0^2,
$
where
$
E_{10}=\sqrt{2|\gamma_1|/k_0|n_{2}^Q|}/W_0=\sqrt{2|\Delta k\gamma_1|/\chi_1\chi_2}/W_0,N_{Q0}=2|\gamma_1|/k_0W_0^2,
$
the dimensionless system can be obtained
\begin{subequations}\label{model-qnp-normal-whole}
\begin{equation}
i\frac{\partial \phi}{\partial z}+\frac{s_1}{2}\frac{\partial^2\phi}{\partial x^2}+\Delta n \phi^*=0,
\end{equation}\begin{equation}
-s_2w_m^2\frac{\partial^2\Delta n}{\partial x^2}+\Delta n-s\phi^2=0,
\end{equation}\end{subequations}
where $w_m=W_{mQ}/W_0$ and $s=\mathrm{sgn}( n_{2}^Q)$. For the spatial case (optical beams), it can be realized that $s=-1$ and $s_2=-1$ by choozing $\Delta k<0$ because $\gamma_1\approx 2\gamma_2>0$ in this case~\cite{Nikolov-pre-2003}; while for the temporal case (optical pulses), both cases of $s=\pm1$ and $s_2=-1$ might be realized~\cite{Buryak-pla-1995} by carefully choozing the signs of $\Delta k$, $\gamma_1$ and $\gamma_2$. Although a set of Equations~(\ref{model-qnp-normal-whole}) are slight different from Eqs.~(\ref{model-whole}) in their two last terms ($\phi^*$ rather than $\phi$ and $\phi^2$ rather than $|\phi|^2$), Eqs.~(\ref{model-qnp-normal-whole}) and (\ref{model-whole}) are the same in the stationary state~\cite{Buryak-pla-1995,Nikolov-pre-2003,Esbensen-pra-2012}. Of course, whether the quadratic nonlinear process can exhibit the transition between the focusing and the defocusing states is still an open equestion, to answer which more careful and deeper investigations should be needed.

\emph{Conclusion}.---
We have discussed the focusing/defocusing property in the nonlocally nonlinear medium with a sine-oscillation response function, and found it depends on generalized degree of nonlocality of the system $\sigma$.
The transition between defocusing and focusing states of the nonlinear refractive index will occur when $\sigma$ goes cross a critical value $\sigma_c$, which is 0.54 for the Gaussian beam. In the case that $s=-1$, the bright and dark soliton solutions are found to exist, respectively, in the range that $\sigma>1.21$ (the focusing side) and in the range that $\sigma<0.31$ (the defocusing side) of the $\sigma$-coordinate. The bright solitons are stable, and the dark soliton solution are unstable due to the existing MI. The theoretical results pave the way to the experimental observation in the nematic liquid crystals with negative dielectric anisotropy and in the quadratic nonlinear media, and are further expected to introduce significant novelties in all-optical devices.

This research was supported by the National Natural Science
 Foundation of China, Grant No.~11474109.


\begin{thebibliography}{99}
\bibitem{beamwidth-definition} By this definition, the beam width $w_r$ of the Gaussian beam $\phi_0(x)$ given by Eq.~(\ref{Gaussian beam}) is just right equal to $w_0$.
\end{thebibliography}
\end{document}